\documentclass[12pt]{article}
\usepackage{epsf}
\def\be{\begin{equation}}
\def\ee{\end{equation}}
\def\bea{\begin{eqnarray}}
\def\eea{\end{eqnarray}}

\def\lsim{\raisebox{-0.6ex}{$\stackrel{\textstyle <}{\sim}$}}

\begin{document}
\begin{titlepage}
\begin{center}
{\Large \bf William I. Fine Theoretical Physics Institute \\
University of Minnesota \\}
\end{center}
\vspace{0.2in}
\begin{flushright}
FTPI-MINN-17/04 \\
UMN-TH-3620/17 \\
March 2017 \\
\end{flushright}
\vspace{0.3in}
\begin{center}
{\Large \bf Partial waves in $B_s^* \bar B_s^*$ production in $e^+e^-$ annihilation near threshold.
\\}
\vspace{0.2in}
{\bf  M.B. Voloshin  \\ }
William I. Fine Theoretical Physics Institute, University of
Minnesota,\\ Minneapolis, MN 55455, USA \\
School of Physics and Astronomy, University of Minnesota, Minneapolis, MN 55455, USA \\ and \\
Institute of Theoretical and Experimental Physics, Moscow, 117218, Russia
\\[0.2in]

\end{center}

\vspace{0.2in}

\begin{abstract}
The production of the vector meson pairs $B_s^* \bar B_s^*$  in the $e^+e^-$ annihilation is generally contributed by two $P$ wave amplitudes with different total spin $S$ of the mesons, $S=0$ and $S=2$, and also by an $F$ wave amplitude. Belle has recently reported a study of the available data at the energy of the $\Upsilon(5S)$ peak in terms of the two $P$ wave contributions only. It is argued here that although at this energy the $F$ wave should be quite suppressed, especially for the strange-bottom vector mesons, the particular studied angular distribution is very sensitive to a presence of even small $F$ wave amplitude due to a significant interference with the dominant $S=2$ $P$ wave. Thus the available data are not conclusive with regards to the partial wave structure of the production amplitude. Additional angular correlations are discussed that can be measured for a full amplitude analysis of the production process.

\end{abstract}
\end{titlepage}

The strong interaction between slow heavy hadrons gives rise to the observed intricate behavior of processes at the onset of open heavy flavors in various channels both in the charmonium-like and bottomonium-like sectors. In particular, the measurements of the production in $e^+e^-$ annihilation of pairs of heavy mesons allow to study the effects of this interaction in the channel with the quantum numbers $J^{PC} = 1^{--}$. The relative yield of the pairs with the pseudoscalar and vector mesons has been discussed since a while ago~\cite{drgg} in terms of the Heavy Quark Spin Symmetry (HQSS). However  in the vicinity of the thresholds a straightforward application of HQSS is not helpful due to the enhanced violation of this symmetry~\cite{mv12}, and apparently additional studies are needed. One important, yet unknown characteristic of the process is the spin-orbital structure of the amplitude for the production of pairs of vector mesons, e.g. $e^+ e^- \to B^* \bar B^*$. Three different partial wave amplitudes are generally allowed: one $P$ wave (denoted here as $P_0$) with zero total spin $S$ of two mesons, $S=0$, as well as a $P$ wave and $F$ wave with $S=2$ ($P_2$ and $F$). At a very low energy above the threshold the $F$ wave is expected to be kinematically suppressed, but can rapidly become prominent as the energy is increased. An experimental study of the partial wave structure for production of strange-bottom meson pairs $B_s^* \bar B_s^*$ is recently reported by Belle~\cite{belle} with bulk of the data collected near the $\Upsilon(5S)$ peak at the c.m. energy between 10863 and 10869\,MeV, i.e. approximately 35 - 40\,MeV above the threshold. Under the assumption of vanishing $F$ amplitude the analysis of the angular distribution of the produced heavy mesons relative to the beam direction gave the fractional contribution of the  $P_0$ - wave to the overall rate the value $r= |P_0|^2/(|P_0|^2 + |P_2|^2) = 0.175 \pm 0.057^{+0.022}_{-0.018}$, which can be compared with the HQSS value~\cite{mv12} $r_0=1/21 = 0.048$. As will be discussed further in this paper, the particular angular distribution used in the analysis of Ref.~\cite{belle} is very sensitive to presence of even a very small $F$ wave, due to its significant interference with the dominant $P_2$ amplitude. Namely, the observed effect attributed to the $P_0$ amplitude can instead be reproduced by a minute $F$ wave contributing only about 0.01 of the total production rate. Due to this observation it becomes of interest to possibly quantify  the expected $F$ wave contribution to the process studied in Ref.~\cite{belle}. It will be argued here that the suppression of the $F$ wave near threshold should be especially effective for the strange $B_s^*$ mesons as opposed to the non-strange ones due to absence of a long-distance interaction induced by the pion exchange. However, even with this suppression it would not be unnatural if the $F$ 
wave amplitude reaches a value that significantly contributes to the measured angular distribution~\cite{belle}. For this reason it appears that further studies of additional angular distributions with future experimental data will be necessary for untangling the spin-orbital structure of the production amplitude. 

An approximate estimate of the significance of the $F$ wave can be done by considering a rescattering between heavy meson pairs that generally results in a $P - F$ mixing.  In the case of non-strange vector meson pairs $B^* \bar B^*$ such rescattering can be analyzed very near the threshold, where the effect of the pion exchange is parametrically enhanced. Using this approach it has been found~\cite{lv13} that the mixing between the $P$ waves, $P_0 - P_2$, can be significant, while the $P - F$ mixing remains very small (less than few percent in the amplitude) in the the range of excitation energy up to 15 - 20\,MeV, where the approach is applicable. This however does not exclude that the $F$ wave rapidly becomes significant at somewhat larger energy, in particular at the energies relevant for the data~\cite{belle}. Furthermore this approach does not apply to the production of the strange $B_s^*$ mesons between which there is no pion exchange. In this case one has to resort to parametric estimates based on the general properties of the scattering amplitudes at a nonzero orbital momentum. 

At the discussed excitation energy of 35 - 40\, MeV the c.m. momentum of each of the $B_s^*$ mesons is $k \approx 450\,$MeV. The lightest hadronic state that can be exchanged between  strange heavy mesons is two pions, so that the longest range $a$ of the strong interaction between them should not exceed $(2 m_\pi)^{-1}$. It is not known how strong this two-pion force between the strange mesons is, or how the strength is distributed in the invariant mass in the $t$-channel. The low invariant mass near the $\pi \pi$ threshold should be somewhat suppressed due to the chiral properties of the pions, so that an estimate $a \, \, \lsim \, \, (300\,{\rm MeV})^{-1}$ appears to be a quite conservative lower estimate for the range of the force. Clearly, the exchange of the lightest meson containing an $s \bar s$ quark pair, the $\eta(548)$, results in a still shorter range of the interaction~\footnote{The interaction between heavy mesons resulting from the $\eta$ exchange has been recently analyzed in Ref.\cite{kr}.}. Another possible rescattering of pairs of non-strange $B^{(*)}$ mesons into $B_s \bar B_s^*$ can proceed due to exchange of a single Kaon and also corresponds to a shorter-range effective interaction.  It can be noted that in the discussed situation the parameter for applicability of an effective radius approximation $k a$ is not small by itself and an evaluation of possible wave(s) with higher angular momentum requires some additional consideration. Indeed, if one treats the motion with a `high' orbital angular momentum $\ell=3$ as semi classical~\cite{ll}, the centrifugal barrier should be taken as $(\ell+1/2)^2/(M r^2)$ with $M \approx 5415\,$MeV being the mass of the $B_s^*$ meson. At the momentum $k$ the particles emerge from under the barrier at the distance 
\be
b = {\ell + 1/2 \over k} \approx  (130 \, {\rm MeV})^{-1}~,
\label{beff}
\ee
which is well beyond the range of the interaction. Thus the `barrier attenuation' of the wave function in the propagation from the interaction distance $a$ to $b$ should be included in the estimate of the suppression of the amplitude with the orbital angular momentum $\ell$. It can be readily seen that the barrier factor depends on the parameter $a/b = ka/(\ell+1/2)$, rather than on $k a$ only. In practice an actual semi classical calculation is not needed and one can arrive at the same result by considering the (normalized at infinity) radial wave function of the free motion with the angular momentum $\ell$~\cite{ll}: 
$R_{k \ell}(r) = 2 k \, j_\ell(k r)  \approx {2 \, k^{\ell +1 }  / (2 \ell +1 )!!} \, r^\ell$ 
and using its value at $r \approx a$. In this way one readily estimates the suppression factor for the $\ell=3$ $F$ wave relative to the $P$ wave:
\be
{F \over P} \sim {(k a)^2 \over 5 \cdot 7 } \, \, \lsim \, \, 0.06~.
\label{fpr}
\ee

Certainly, in the estimate in Eq.(\ref{fpr}) no information about the dynamics is taken into account besides the general notion of the range $a$ of the strong interaction, and actual details of the production mechanism can result in a significant modification of the relative strength of the partial waves. For this reason a further study of the discussed process is needed. It is important however, that as small as the $F$ wave amplitude in the estimate (\ref{fpr}) is, it can nevertheless be well visible in the angular distributions.  In order to argue this point one can write the production amplitude in the c.m. system of the $e^+e^-$ beams in terms of the properly normalized three discussed amplitudes as
\bea
&&A \left ( e^+ e^- \to B_s^* \bar B_s^* \right ) = P_0 \, (\vec j \cdot \vec n) \, (\vec \epsilon_1 \cdot \vec \epsilon_2) + {3 \over 2 \, \sqrt{5}} \, P_2 \, j_i n_k \, \left [ \epsilon_{1i} \epsilon_{2k} + \epsilon_{2i} \epsilon_{1k} - {2 \over 3} \, \delta_{ik} \, (\vec \epsilon_1 \cdot \vec \epsilon_2) \right ] \nonumber \\
&& + { \sqrt{15} \over \sqrt{2}} \, F \, j_i \left ( n_i n_k n_l - {1 \over 5} \, \delta_{ik} \, n_l - {1 \over 5} \, \delta_{il} \, n_k  - {1 \over 5} \, \delta_{kl} \, n_i \right ) \epsilon_{1k} \epsilon_{2l}~,
\label{amp}
\eea
where $\vec \epsilon_1$ and $\vec \epsilon_2$ are the polarization amplitudes of the two vector mesons, $\vec n$ is a unit vector in the direction of their relative motion, and $\vec j$ is proportional to the electromagnetic current of the electron and positron $\vec j = {\rm const} \cdot (\bar e \vec \gamma e)$ where the constant can be chosen in such a way that the total cross section is expressed in terms of the partial amplitudes with no extra coefficients:
\be
\sigma \left ( e^+ e^- \to B_s^* \bar B_s^* \right ) = |P_0|^2 + |P_2|^2 + |F|^2~.
\label{xsec}
\ee

\begin{figure}[ht]
\begin{center}
 \leavevmode
    \epsfxsize=15cm
    \epsfbox{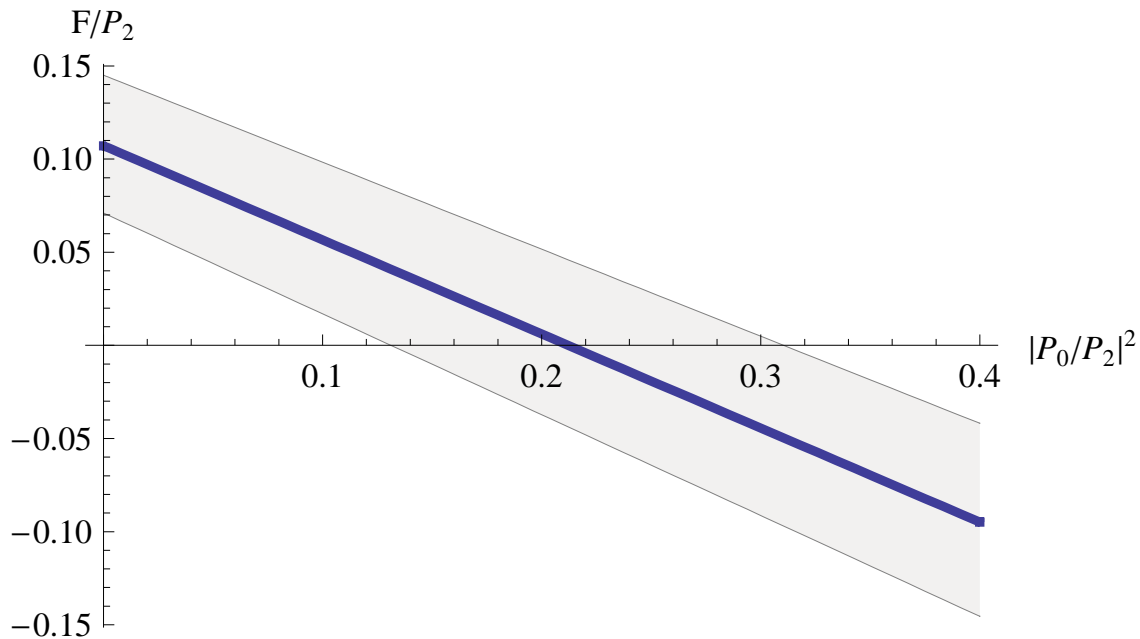}
    \caption{The values of the quadratic ratio $|P_0/P_2|^2$ and the linear one $F/P_2$ corresponding to the result of the angular analysis in Ref.~\cite{belle}, assuming real $F/P_2$. The shaded band is the one sigma error corridor with the statistical and systematic errors combined in quadrature. }  
\end{center}
\end{figure}

It is quite straightforward to derive from Eq.(\ref{amp}) the distribution of the production rate over the angle $\theta_{e B}$ between the direction of the beams and that of the produced mesons. Introducing the notation $u= \cos\theta_{eB}$ one finds the expression for the angular correlation studied by Belle~\cite{belle}
\bea
&&{d \sigma \over du } = {3 \over 4} \, |P_0|^2 \, (1- u^2) + {21 \over 40} \, |P_2|^2 \, \left ( 1 -{1 \over 7} \, u^2 \right )  \nonumber \\ 
&& + {3 \over 5} \, |F|^2 \left ( 1 - {1 \over 2} u^2 \right ) + {3 \sqrt{3} \over 10 \sqrt{2} } \, {\rm Re} (P_2 \, F^*) \, (1 - 3 u^2)~.
\label{du}
\eea
As it should be expected, the $S=0$ and $S=2$ amplitudes do not interfere in this distribution, while the interference between the two amplitudes with $S=2$ is potentially very significant. Indeed, the relative coefficient of $u^2$ in the angular distribution depends on three dimensionless ratios of the partial wave amplitudes: $|P_0/P_2|^2$, Re$(F/P_2)$ and Im$(F/P_2)$, so that any measured value would generally describe a two-dimensional surface in the space of these three variables.  With the general understanding that the $F$ wave amplitude is small as expected from Eq.(\ref{fpr}), the variable Im$(F/P_2)$ is small and enters quadratically and can thus be neglected.  However the relatively real part, Re$(F/P_2)$, enters linearly in the interference term in Eq.(\ref{du}) with a numerically large coefficient. (In fact the value of the latter coefficient is the maximal one allowed by the positivity of the rate.) As a result a measurement of the relative coefficient of $u^2$ may not yield a conclusive outcome for the partial wave composition of the production amplitude, even if the $F$ wave contribution is as small as suggested by the estimate in Eq.(\ref{fpr}). This observation is illustrated in Fig.~1, where the solid line corresponds to the values of the ratios $|P_0/P_2|^2$ and $F/P_2$ that would describe the central value of the result of the angular analysis of Ref.~\cite{belle}. In particular the whole observed effect attributed to the contribution of the $P_0$ amplitude can be rather described by a small $F$ wave with a magnitude in the same ballpark as in Eq.(\ref{fpr}). 

Additional angular distributions that may be used for further studies of the discussed amplitude may involve detection of the direction of one photon emerging from the decay $B_s^* \to B_s \gamma$ or $\bar B_s^* \to \bar B_s \gamma$, or detection of both photons, since the direction of the photon emission is correlated with the spin of the heavy vector meson. In the setting where only one photon is detected (from either of the mesons in the pair) one ca define, in addition to the previously considered angle $\theta_{eB}$, also the angle between the photon and the beams, $\theta_{e \gamma}$,  and the angle $\theta_{\gamma B}$ between the photon and the direction of the motion of the heavy mesons. Using the standard expression for the amplitude of the electromagnetic decay of the vector meson,
\be
A(B^* \to B \gamma) = g ([\vec q \times \vec \epsilon_\gamma] \cdot \vec \epsilon_{B^*})
\label{ag}
\ee
with $\vec q$ being the momentum of the photon and $\vec \epsilon_\gamma$ its polarization amplitude, and also introducing the notation for the cosines of the extra angles: $v = \cos \theta_{e \gamma}$ and $w = \cos \theta_{\gamma B}$, one can find the double differential angular distribution in the form
\bea
&&d \sigma  \propto {3 \over 4} \, |P_0|^2 \, (1-u^2) + {3 \over 160} \, |P_2|^2 \, (29-2 u^2 + 9 v^2 - 12 w^2 - 6 u v w) + 
\nonumber \\ 
&&{9 \over 80} \, |F|^2 \, (6 - 3 u^2 + v^2 - 3 w^2 - 4 u v w + 5 u^2 w^2) + \nonumber \\
&& {3 \over 4 \, \sqrt{5}} \, {\rm Re} (P_0 P_2^*) \, (1-u^2 -3 w^2 + 3 u v w) + {3 \, \sqrt{3} \over 4 \, \sqrt{10}} \, {\rm Re} (P_0 F^*) \, (1-u^2 - 3 w^2 - 2 u v w + 5 u^2 w^2) \nonumber \\
&&+ {3 \, \sqrt{3} \over 40 \, \sqrt{2}} \, {\rm Re} (P_2 F^*) \, (7 - 16 u^2 - 3 v^2 - 6 w^2 +7 u v w +5 u^2 w^2)~.
\label{g1dd}
\eea
It should be noticed that this distribution does not depend on the flavor identification of the $B^*$ meson  (i.e. $B^*$ or $\bar B^*$). Formally, this corresponds to the symmetry under the change of the sign of the `$B$' direction ($u \to -u$, $w \to -w$).

If averaged over the direction of the photon (this correponds to the replacement $v^2 \to 1/3$, $w^2 \to 1/3$, $v w \to u/3$), the distribution in the remaining variable $u$ reproduces the one in Eq.(\ref{du}). On the other hand if the direction of the $B$ mesons is averaged out (this corresponds to the replacement $u^2 \to 1/3$, $w^2 \to 1/3$, $u w \to v/3$, $u^2 w^2 \to 1/15 + 2 v^2/15$), the distribution over the angle between the photon and the electron-positron beams is given by
\be
{d \sigma \over dv} = {1 \over 2} \, |P_0|^2 + {1 \over 160} \, |P_2|^2 \, ( 73+ 21 v^2) + { 3 \over 80} \, |F|^2 \, (13 + v^2) - {1 \over 4 \, \sqrt{5}} \, {\rm Re} (P_0 P_2^*) \, (1 - 3 v^2)~,
\label{dv}
\ee
in which case there is no interference of the waves with different $\ell$, i.e. between the $F$ and either of the $P$ waves.

For the setting, where the direction of both photons from the decays of $B^*$ and $\bar B^*$ is measured, one can introduce the notation 
$v = \cos \theta_{e \gamma1}$, $y=\cos \theta_{e \gamma2}$, and $z=\cos \theta_{\gamma1 \, \gamma2}$. Then the double differential distribution is given by the expression
\bea
\label{g2dd}
&&d \sigma \propto {3 \over 8} \, |P_0|^2 \, (1+z^2) + {3 \over 160} \, |P_2|^2 (22 + 6 v^2 + 6 y^2 +z^2+ 3 v y z) + \\ \nonumber&&{9 \over 560} \, |F|^2 \, (29 + 2 v^2 + 2 y^2 + 2 z^2 + v y z)  
-{3 \over 8 \, \sqrt{5}} \, {\rm Re} (P_0 P_2^*) \, (2 - 3 v^2 - 3 y^2 - z^2 +3 v y z) ~,
\eea
and there is no interference between the $P$ and $F$ wave the amplitudes. 

If the beam direction is averaged out, the distribution in the angle between the photons reads as
\be
{d \sigma \over dz}= {3 \over 8} \, |A_0|^2 \, (1+z^2) + {3 \over 80} \, (|A_1|^2 + |A_3|^2) \, (13+z^2) ~,
\label{dz}
\ee
(the interference between $P_0$ and $P_2$ also vanishes here). If the distribution in Eq.(\ref{g2dd}) is averaged over the direction of one of the photons ($y^2 \to 1/3,\, z^2 \to 1/3, \, v y z \to v^2/3$)  the resulting distribution in the remaining variable $v$ reproduces the one given by Eq.(\ref{dv}).

To summarize. The $F$ wave amplitude in the production of the $B_s^* \bar B_s^*$ strange-bottom meson pairs should is expected to be quite small, especially given that the range of the strong interaction between these mesons should be shorter than for the non-strange ones. However even a small $F$ wave amplitude may significantly affect through the interference the angular distributions in the production process, in particular the one recently measured by Belle~\cite{belle}. Thus a full study of the partial waves in the $e^+e^-$ annihilation to pairs of heavy vector mesons would require measurement of additional angular correlations, such as the double differential distributions in described by the formulas (\ref{g1dd}) and (\ref{g2dd}), or at least their simplified versions in Eqs.(\ref{dv}) and (\ref{dz}).  
 
I acknowledge stimulating discussions with Alexander Bondar.
This work is supported in part by U.S. Department of Energy Grant No.\ DE-SC0011842.

\end{document}